\documentclass[12pt]{article}

\usepackage{epsfig}
\usepackage{cite}
\usepackage{latexsym} 
\usepackage{a4}
\usepackage{amsfonts}
 \newcommand{\plabel}{\label}
\def\ul{\underline}
\newcommand{\oB}{|_{\partial{\cal M}}} 

\begin{document}

\begin{titlepage} \renewcommand{\thefootnote}{\fnsymbol{footnote}}

\hfill{}TUW--00--18 \\

%\hfill{}DRAFT \today \\

{\par\centering \vspace{1cm}\par}

{\par\centering \textbf{\large 
Renormalizability of the open 
string sigma model and emergence of $D$-branes
}\large \par}

{\par\centering \vspace{1.0cm} \vfill
\renewcommand{\baselinestretch}{1}\par}

{\par\centering \textbf{W.\ Kummer\( ^{1} \)\footnotemark[1] 
and D.V.\ Vassilevich\( ^{2} \)\footnotemark[2] \footnotemark[3]{}} 
\vspace{7ex}\par}

{\par\centering \( ^{1} \)Institut f\"{u}r Theoretische Physik, Technische
Universit\"{a}t Wien \\
 Wiedner Hauptstr. 8--10, A-1040 Wien \\
 Austria \vspace{2ex}\par}

{\par\centering \( ^{2} \)Institut f\"{u}r Theoretische Physik, Universit\"{a}t
Leipzig,\\
 Augustusplatz 10, D-04109 Leipzig,\\
 Germany\par}

{\par\centering \footnotetext[1]{e-mail: \texttt{wkummer@tph.tuwien.ac.at}} \footnotetext[2]{e-mail:
\texttt{Dmitri.Vassilevich@itp.uni-leipzig.de}} \footnotetext[3]{On leave from
Department of Theoretical Physics, St. Petersburg University, 198904 St. Petersburg,
Russia}\par}

\vfill

\begin{abstract}
Rederiving the one-loop divergences for the most general 
coupling of the open string sigma model by the heat kernel 
technique, we distinguish the classical 
background field from the mean field of the effective action. 
The latter is arbitrary, i.e.\  
does not fulfil the boundary conditions. As a consequence a 
new divergent counter term strongly suggests  the introduction of another 
external one-form field (beside the usual gauge field), 
coupled to the normal derivative at the boundary. Actually 
such a field has been proposed in the literature for 
different reasons, but its full impact never seems to have 
thoroughly investigated before. The beta function for the 
resulting renormalizable model is calculated and the 
consequences are discussed, including the ones for the 
Born-Infeld action. The most exciting property of the new
coupling is that it enters the coefficient in front of the normal
derivative in Neumann boundary conditions. For certain values of the
background fields this coefficient vanishes, leading to Dirichlet
boundary conditions. This provides 
 a natural mechanism for the 
emergence of D-branes. 
\end{abstract}
\end{titlepage}

\vfill

\section{Introduction}

An essential ingredient for the proper formulation of systems consisting of
strings and (D-) branes is the Born-Infeld (BI) action
which originally had been introduced in string theory
by Fradkin and Tseytlin \cite{FT85,T86}. Other pioneering works
\cite{ACNY87,Callan88} re-derived it from the condition that
the beta-function for the string vanishes when the string is embedded in
an external gauge field $A_\mu (X)$, where $X^\mu$ is the 
target space coordinate. Previously the beta-function method was
successfully applied to closed strings \cite{CFMP85}.
 From the extensive literature on the application of
this action in brane theory and its vast number of consequences 
we only refer to a review \cite{T99}. 

The bulk contribution of the sigma 
model from the usual string arguments also contains the 
two-form gauge field $B$ and two scalar fields ($\Phi, T$) one 
of which $(\Phi)$ is coupled to the world volume
Ricci scalar on the 
(compact) string manifold ${\cal M}_2 ={\cal M}$. In the presence of a 
boundary $\partial {\cal M}$, there beside the one-form gauge field 
$A$ two similar scalar fields ($\hat \Phi, \hat T$) may 
appear, of which $\hat \Phi$ is coupled to extrinsic 
curvature. It has been suggested 
\cite{dornotto86,DLP89,Leigh89,behrndtdorn92,dornotto96}
(e.g.\ from arguments like 
duality) to introduce also  another one-form field $V_\mu$ 
coupled to the normal derivative of $X^\mu$ on
 $\partial {\cal M}$, however we are not aware 
of a systematic treatment of the problems including all 
consequences of that extension. It will be the main issue of 
our present work to show that renormalizability at the 
boundary strongly suggests the inclusion of $V$. As we explain below,
this means an extension of the approach \cite{Callan88}
to achieve full renormalization of the theory. 
At the same time we discover a natural mechanism for the
emergence of $D$-branes.

We have found the heat kernel technique \cite{Gilkey} especially 
convenient for a treatment of the one-loop quantum effects, 
because also for  problems with nontrivial boundary many 
mathematical rigorous results are available which may be 
used in the present context. Our argument 
relies upon the fact that for the one-particle irreducible
(1-p-i) loop 
integral the external lines refer to the ``mean field'' 
$\bar X$, which is arbitrary, and, in particular, does {\em 
not} fulfil the boundary conditions. 

For this reason in 
Sect.\ 2 the relation of $\bar X$ to the background field, 
appearing in the path integral of the generating functional 
of Green functions is clarified.  Instead of doing the 
Legendre transformation in the path integral -- a procedure
which may introduce certain difficulties in the presence
of boundaries -- we show how to construct directly the functional
which generates the one-loop 1-p-i 
diagrams. At this order, this functional is equivalent
to just one diagram with the propagator modified by all
possible vertex insertions. The sketch of a 
 more formal derivation of the same result where the Legendre
transform is used, for a general action with boundary 
is presented in Appendix A. 

In Sect.\ 3 we return to the determination of the divergent 
contribution of the one-loop corrections for the usual 
sigma-model with boundary (without second 1-form field $V$), 
using the powerful heat kernel technique.  Some
mathematic issues as, for example, strong ellipticity
of the boundary value problem are clarified. We rederive the divergent
part of the one-loop effective action and, as a by-product, 
we generalize and extend
the results existing in the literature 
\cite{dornotto86,Callan88,Osborn91}. On the basis of the analysis in
the previous Section
we suggest an extension
of the method of Callan and coworkers \cite{Callan88} which
amounts to the inclusion of the $V$-coupling at the boundary.

In Section 4 we investigate the consequences of introducing the vector
field $V_\mu$, coupled to $\partial_N X^\mu$ in the action, right
from the start. For simplicity we drop in that Section all dimensional
couplings. We analyse the corresponding boundary value problem
and define modified boundary conditions.
An important point is to achieve
hermiticity of the relevant operator. The $\beta$-functions are
explicitly calculated to the leading order of $V$. We show
that the expansion in the $V$-field generates derivative
corrections to the Born-Infeld action with respect to $B_{\mu\nu}$.
If $V$ is no longer small we observe a mechanism for
spontaneous $D$-brane creation, which seems to be the most remarkable
property of the proposed model. Generically, a $D$-brane is
separated 
by a non-perturbative region populated by
tachyonic modes.

%%%%%%%%%%%%%%%%%%%%%%%%%%%%%%%%%%%%%%%%
\section{Effective action in the presence of boundaries}
\label{effective}

The basis of the computation of quantum corrections, 
including renormalization and beta-functions, are 1-p-i 
vertex functions. Generally these vertices are obtained by 
functional differentiation with respect to an (arbitrary) 
``mean field'' $\bar X$ from the effective action $W(\bar 
X)$. In order to see that this mean field in $W$ is different from 
a classical background field we consider as an illustration 
a theory described by the classical action in Euclidean in 
$d=2$ 
\begin{equation}
S=\int_{\cal M} \left( -(\partial X)^2 +w(X)\right) +
\int_{\partial {\cal M}}( - V(X)\partial_N X +
\hat w(X) ) \,,\label{Sclassic}
\end{equation}
where $w(X)$ and $\hat w(X)$ are some ``potential'' terms
which may contain linear first derivatives (in the case of $\hat w$ these  can
be tangential derivatives to the boundary $\partial {\cal 
M}$ only).
$\partial_N$ is the partial derivative with respect to
the inward pointing unit vector on $\partial {\cal M}$. 
In anticipation of the situation for the string, to be discussed below, 
we also allow for a 
boundary  term $V\partial_NX$ with  normal derivative.
Our strategy is to derive the effective action as the 
inverse of the full one-loop propagator in the presence of a 
boundary. 

The action (\ref{Sclassic}) is split into two parts:
\begin{equation}
S=S_{{\mbox {\scriptsize {prop}}}}+S_{{\mbox {\scriptsize {int}}}}
\,,\qquad S_{{\mbox {\scriptsize {prop}}}}=
-\int_{\cal M} (\partial X)^2 \,.
\label{Ssplit}
\end{equation}
$S_{{\mbox {\scriptsize {prop}}}}$ will be used to define propagator,
while $S_{{\mbox {\scriptsize {int}}}}$ determines the vertices.
Usually the quadratic part of $w$ is included in 
$S_{{\mbox {\scriptsize {prop}}}}$. However, since we are going
to sum anyway all possible insertions we are allowed to include
the quadratic terms in $S_{{\mbox {\scriptsize {int}}}}$.
The propagator following from $S_{{\mbox {\scriptsize {prop}}}}$
above is 
\begin{equation}
G_0(x,y)=(-\partial^2)_{x,y} \label{G0}
\end{equation}
which is subject to boundary conditions defined by the
boundary part of $S_{{\mbox {\scriptsize {prop}}}}$.
These can be either of  Dirichlet 
\begin{equation}
G_0(x,y)|_{x\in {\partial {\cal M}}}=0 \label{DirBC}
\end{equation}
or of Neumann type 
\begin{equation}
\partial_N^xG_0(x,y)|_{x\in {\partial {\cal M}}}=0\; . \label{NeuBC}
\end{equation}
Theories described by these two types
of the boundary conditions are different already at the
classical level. Here we consider Neumann conditions (\ref{NeuBC})
only, because their specific feature will be essential for 
our work below. 

For our purposes it will be enough to treat 
1-p-i Feynman diagrams 
with $n$ external lines. In coordinate space any such diagram
is given by an $n$-point function ${\cal W}_{(n)}(x_1,\dots ,x_n)$.
The renormalization problem refers to
{\it amputated} diagrams. This means that the external
lines just label arguments in the  vertices. Thus to obtain the 
one-loop effective action the mean field
$\bar X(x_i)$ may be simply attached at the vertices. 
Summing all such diagrams 
\begin{equation}
W(\bar X)=\sum_{(n)} \int dx_1 \dots \int dx_n 
{\cal W}_{(n)}(x_1,\dots ,x_n) \bar X(x_1) \dots \bar X(x_n)
\label{effact}
\end{equation}
the effective action $W$ is obtained to one-loop. Clearly 
by functional differentiation with respect to the mean field $\bar X$
the individual 1-p-i diagrams are recovered. 

The correct combinatorial coefficients of the vertex insertions in
the propagator can be obtained by taking, as a first step,  
$S_{{\mbox {\scriptsize {int}}}}(\bar X +\xi )$,  keeping 
only the second order terms in $\xi$:
\begin{equation}
S^{(2)}_{{\mbox {\scriptsize {int}}}}=
\frac 12 \int_{\cal M} \xi w''(\bar X) \xi 
+\int_{\partial {\cal M}} (-V'(\bar X) \xi\partial_N \xi
-\frac 12 \xi V''(\bar X)\xi \partial_N \bar X
+\frac 12 \xi \hat w''(\bar X) \xi ) 
\label{S2int}
\end{equation}
Then the propagator $G_1(x,y)$ generated by all
insertions of boundary vertices in $G_0(x,y)$ becomes 
\begin{eqnarray}
&&G_1(x,y)=\sum_{k=0}^\infty \int_{\partial {\cal M}}dx_1 \dots
\int_{\partial {\cal M}}dx_k
(-1)^k G_0(x,x_1)U(x_1)G(x_1,x_2)\dots
U(x_k)G_0(x_k,y)\; , \nonumber \\
&&U(x)=-\overleftarrow{\partial}_NV'(\bar X)
-V'(\bar X)\overrightarrow{\partial}_N -\frac 12 V''(\bar X)
(\partial_N\bar X)
+\frac 12 \hat w''(\bar X) \; .
\label{G0G1}
\end{eqnarray}
It should be noted that the normal derivative appears twice in the
boundary vertex operator $U$. This happens because
on the boundary $\xi$ and $\partial_N\xi$ should be
considered as independent.  All correlation functions of the
field $\xi$ should follow by variation of the path
integral containing $\int_{\cal M} (j\xi )$ with
respect to the source $j$. If one needs correlators
containing $\xi (x)$, the source $j$ should also include a 
term behaving like a $\delta$-function at the boundary. 
To obtain derivatives of
$\xi$ that part of 
the source should contain also a derivative of the
$\delta$-function and an integration by parts should be performed.
This does not work, however, if one needs {\it normal}
derivatives of $\xi$ on the boundary. Then that integration
by parts of terms like $\int_{\cal M}\xi \partial_N 
\delta (\partial{\cal M})$  would lead to
a meaningless expression containing a factor 
$(\delta (\partial{\cal M}))^2$ in the integral 
(with one delta function coming from the integrand, and the other
describing the surface contribution). To overcome this difficulty we 
introduce an independent field 
$\chi =\partial_N \xi$ on 
the boundary with the propagators $G_0^{\chi \xi}(x,y)=
\partial^x_N G_0(x,y)$ and $G_0^{\chi \chi}(x,y)=
\partial^x_N \partial_N^yG_0(x,y)$. Note, that one cannot put this
propagator to zero if it appears under an integral over
$x$ because of the singularity in $G_0$ for coinciding
arguments. This leads to the effective doubling of the
normal derivative term in $U$ in (\ref{G0G1}) because one
should take into account both terms: $\dots G_0^{\dots \chi}
(x_{n-1},x_n)(-V')G_0^{\xi\dots}(x_{n},x_{n+1})\dots$
and $\dots G_0^{\dots \xi}
(x_{n-1},x_n)(-V')G_0^{\chi\dots}(x_{n},x_{n+1})\dots$.

For the next step we note that $G_1(x,y)$ evidently satisfies the Dyson
equation:
\begin{equation}
G_1(x,y)=G_0(x,y)-\int_{\partial {\cal M}} dz G_0(x,z)U(z)
G_1(z,y) \label{inteq}
\end{equation}
In order to find out whether the correct boundary conditions are
satisfied by $G_1(x,y)$ we let
$\partial_N^x$ act on both sides of the equation (\ref{inteq})
and put $x$ onto the boundary: 
\begin{equation}
\partial_N^x G_1(x,y) |_{x\in {\partial{\cal M}}}=
-\lim_{x\to {\partial{\cal M}}} 
\int_{\partial {\cal M}} dz\partial_N^x   G_0(x,z)U(z)
G_1(z,y) \label{1857}
\end{equation}
To calculate the right hand side of (\ref{1857}) we 
restrict ourselves to the case of the half-plane,
 ${\cal M}={\bf R}\times {\bf R}_+$. As we will see below,
only the leading short-distance singularity of the propagator $G_0(x,z)$
contributes to (\ref{1857}). At short distances all two
dimensional manifolds with boundary, as far as the leading singularity is
concerned, can be treated  as a
half-plane. This restriction allows to take the 2d propagator $G_0$ 
simply as the one with a ``mirror charge''
\begin{equation}
G_0(x,z)=-\frac 1{4\pi} (\ln |x-z|^2 + \ln |x-z^*|^2 ) \label{1911}
\end{equation}
where $z^*=(z^1,-z^2)$. For symmetry reasons the contribution 
to (\ref{1857}) of the 
first term in $U$ of  (\ref{G0G1}) vanishes. For the rest
of  $U$ we have
\begin{eqnarray}
&&\partial_N^x G_1(x,y) |_{x\in {\partial{\cal M}}}= \frac 1{4\pi}
\lim_{x^2\to +0} \int_{\partial {\cal M}} dz
\frac {4x^2}{(x^2)^2 +(x^1-z^1)^2} \times \nonumber \\
&&\qquad\qquad\qquad\qquad \times \left(
-V'(\bar X)\overrightarrow{\partial}_N^z -\frac 12 V''(\bar X)
(\partial_N\bar X)
+\frac 12 \hat w''(\bar X) \right) G(z,y) \nonumber \\
&&\qquad\qquad = \left(
-V'(\bar X)\overrightarrow{\partial}_N^x -\frac 12 V''(\bar X)
(\partial_N\bar X)
+\frac 12 \hat w''(\bar X) \right) G(x,y)
|_{x\in {\partial{\cal M}}} \label{1402}
\end{eqnarray}
where the identity
\begin{equation}
\lim_{\epsilon\to 0}\frac \epsilon{\epsilon^2 +z^2} =
\pi \delta (z) \label{limdel}
\end{equation}
has been used. It is straightforward to check that also 
\begin{equation}
-\partial^2 G_1(x,y)=\delta (x-y)\; . \qquad x,y\not\in 
{\partial{\cal M}} \label{1427}
\end{equation}

Thus $G_1$ is in agreement with the boundary term in 
(\ref{S2int}) and it represents a proper solution of $- 
\partial^2$. Finally 
 we sum  all the bulk insertions of (\ref{S2int}) to the propagator
$G_1(x,y)$. Clearly, the full propagator satisfies
\begin{equation}
G(x,y)=G_1(x,y)-\int_{\cal M} dz G_1(x,z)\frac 12 w''(\bar X(z))
G(z,y) \, .
\label{G1G}
\end{equation}
Since $w''(\bar X(z))$ is smooth on $\partial {\cal M}$ 
 the boundary
conditions remain valid for $G(x,y)$ as well as for $G_1(x,y)$.
The bulk equation (\ref{1427}) is modified by the term 
proportional to $w''$ 
\begin{eqnarray}
&&\partial^2_x G(x,y)=-\delta (x,y)-\int_{\cal M} dz \partial_x^2 G_1(x,z)
\frac 12 w''(\bar X(z)) G(z,y) \nonumber \\
&&\qquad\qquad\ =-\delta (x,y) +\frac 12  w''(\bar X(x)) G(x,y)
\label{eqmo}
\end{eqnarray}
for $x,y\not\in {\partial{\cal M}}$ so that $G$ contains the 
``mass'' term in the proper way. 

Thus we have proved that,  after summing  all insertions in the
propagator $G_0(x,y)$ which are relevant for calculation of the
one-loop effective action,  one arrives at  a propagator 
\begin{equation}
G(x,y)=\left( -\partial^2 +\frac 12 w''(\bar X)\right)^{-1}_{x,y} 
\label{fullprop}
\end{equation}
defined on the space of fields satisfying  generalized
Neumann boundary condition
\begin{equation}
\left( -\partial_N -V'(\bar X)\partial_N  -\frac 12 V''(\bar X)
(\partial_N\bar X)
+\frac 12 \hat w''(\bar X) \right) \xi |_{\partial {\cal 
M}}=0\; .
\label{genNBC}
\end{equation}
>From this we can conclude that the effective 
action becomes 
\begin{equation}
W(\bar X)=\frac 12 \ln \det 
\left( -\partial^2 +\frac 12 w''(\bar X)\right) \; ,
\label{ea1504}
\end{equation}
the ``naive'' result one would expect based upon  experience
for manifolds without boundaries. 

Though our calculations were valid, strictly speaking, only for
flat geometries, the generalization to curved manifolds with curved
boundaries is straightforward,  but technically involved. Another
subtle point is the symmetry of the propagator $G(x,y)$ which is
equivalent to the hermiticity of the operator in (\ref{eqmo})
when (\ref{genNBC}) holds. This problem will be discussed in
sec. \ref{bcsection}.

In a suitable regularization the divergent part 
$W_{{\mbox {\scriptsize {div}}}}$ of $W(\bar X)$
defines the counterterms which should introduced in the
theory for renormalization. {\it Any} local invariant
(bulk or boundary) appearing in $W_{{\mbox {\scriptsize {div}}}}$
corresponds to certain divergent diagrams just by definition
of the effective action. It is clear therefore that we
are not allowed to impose any restrictions, as e.g.
boundary conditions, on the mean field $\bar X$. 
If  $W(\bar X)$ as a function
of some {\it restricted} $\bar X$ is made finite, 
we cannot guarantee that
the {\it full} theory will become finite as well.
Of course, if we contract the mean field with {\it
external} lines of a digram, the contributions proportional
to the boundary operator will vanish. Therefore, to make the
theory finite at one loop it is enough to remove divergences
which are not annihilated by the boundary operator (as done
in \cite{Callan88}). However, if the mean field represents
{\it internal} lines of a diagram, there could appear problems
with renormalizability due to interference of singularities.
It is not clear, to which order of perturbation theory
a renormalization scheme with boundary conditions on the
mean field could be extended (in certain models such an extension 
has been achieved up to two loops \cite{McO2}). 

The necessity to introduce independent counterterms was recognized
long ago by Symanzik \cite{Symanzik}. Also  related 
work \cite{Wipf94} should be mentioned where $\beta$-functions with
Dirichlet boundary conditions were considered. Although in that
paper the boundary value of the mean field had been fixed at
some intermediate steps, in the calculations of the $\beta$-functions
it was a free parameter generating all counter-terms.

In the analysis 
of the renormalization
of the bosonic sigma model without boundaries 
in ref. \cite{Howe88} 
a ``shift'' symmetry $\delta \bar X=-\delta\xi=\eta$
was introduced. This symmetry has been proposed also
in the case of boundaries \cite{Osborn91} where
boundary conditions on $\bar X$ were adapted by that
symmetry. In our
view, although this helps to select appropriate
boundary conditions, it nevertheless is not
relevant for the question 
whether boundary conditions for $\bar X$ should be imposed at all.

In Appendix A a more formal derivation of the main result of the
present Section
for a general action with boundary condition is sketched, 
because this may indicate a starting point for 
considerations valid beyond the one-loop argument given 
here.

%%%%%%%%%%%%%%%%%%%%%%%%%%%%%%%%%%%%%%%%%%%%%%%%%%%%
\section{Conventional sigma model }

\subsection{Background field formalism}

The action for the sigma model  is usually written as 
\begin{eqnarray}
&&S= \int _{\cal M}d^{2}z\sqrt{h}
\left(\frac{1}{2} G_{\mu\nu} h^{ab}
\partial _{a}X^{\mu }\partial _{b}X^{\nu }
+\frac{1}{2\sqrt{h}}\epsilon^{ab}B_{\mu\nu}
\partial _{a}X^{\mu }\partial _{b}X^{\nu } +\frac 12 \Phi R +T \right)
\nonumber \\
&&\qquad+\int _{\partial {\cal M}}d\tau \left(
 A_{\mu }\partial _{\tau }
X^{\mu } +\hat\Phi k +\hat{T} 
\right) \label{act}
\end{eqnarray}

On the world sheet we consider Euclidean signature space coordinates 
$x^a$ and the metric $h_{ab}$ ($R$ is the related Ricci-scalar 
on ${\cal M}$, $k$ the extrinsic curvature on the boundary), 
$X^\mu (x)$ are target space coordinates with metric 
$G_{\mu\nu}$. Further external fields are the two-form gauge 
field $B_{\mu\nu}$ and scalar fields $\Phi$ and $T$ on ${\cal M}$ 
and the usual gauge field $A_\mu$ together with the scalars 
$\hat\Phi$ and $\hat T$ at the boundary with tangential 
variable $\tau$. The string tension $\alpha'$ has been
absorbed in the external fields.

It should be emphasized that in our convention during the 
transition to an imaginary time coordinate we have also 
rotated $A_\mu \to i\, A_\mu$ and $B_{\mu\nu}\to iB_{\mu\nu}$. 
This rule of continuation has 
the important advantage to provide a well-posed spectral 
problem in Euclidean space with the {\em same} conjugation 
properties of all fields. Of course, that rule is not 
mandatory. The action with imaginary
coefficients can be employed as well
 at the expense of introducing a more complicated
conjugation operation \cite{Osborn91}. Since hermiticity
properties will play an important role in our study, we prefer
to have as simple conjugation rules as possible. Actually,
the fact that parity odd fields may acquire an imaginary factor 
in Euclidean space has been observed long ago \cite{AB} in 
the context of chiral theories. It is easy to see that the 
field $A_\mu$ is of odd parity from the world volume point 
of view. Indeed, the $A$-term in (\ref{act}) can be rewritten as 
$\int_M \partial_a (\varepsilon^{ab} A_\mu \partial_b 
X^\mu)$.  Here $A_\mu$ couples to the parity-odd quantity 
$\varepsilon^{ab}$. 
Whenever it is possible to compare our results to those of previous related
papers \cite{ACNY87,Callan88,Osborn91},
they are compatible after the replacement \( A\rightarrow iA,
B\rightarrow iB \).

As shown in the previous Section the expansion of 
(\ref{act}) into a classical background solution and 
(geodesic) fluctuations $\xi$ allows that background field  
to be identified directly with the (arbitrary) mean field 
$\bar X$, when the one-loop contribution to the effective 
action is sought. Proceeding in the standard way 
\cite{braaten85} produces
the quadratic action
\begin{eqnarray}
&&S_2=\frac 12 \int _{\cal M}d^{2}z\sqrt{h} \left(
G_{\mu\nu}(\bar X) h^{ab}
\nabla_{a}\xi^{\mu }\nabla_{b}\xi^{\nu }
-{\cal R}_{\mu\nu\rho\sigma}(\partial_a\bar X^\mu )
(\partial^a\bar X^\sigma )\xi^\nu\xi^\rho
\right.\nonumber \\
&& \qquad -\frac 12 (\partial_a\bar X^\rho )
(\partial_b\bar X^\sigma )\epsilon^{ab}\xi^\mu\xi^\nu
D_\mu H_{\nu\rho\sigma} 
 +\frac 14 (\partial_a\bar X^\nu )
(\partial^a\bar X^\rho ) H_{\mu\nu\lambda}{H^\lambda}_{\rho\sigma}
\xi^\mu\xi^\sigma \nonumber \\
&& \qquad \left. +\frac 12 (D_\mu D_\nu \Phi (\bar X) )\xi^\mu\xi^\nu R
+(D_\mu D_\nu T (\bar X) )\xi^\mu\xi^\nu \right) \label{S2} \\
&&\qquad +\frac 12 \int _{\partial {\cal M}}d\tau \left(
\xi^\nu D_\tau \xi^\mu (B_{\mu\nu}-F_{\mu\nu}) +
(\partial_\tau \bar X^\rho ) D_\mu (F_{\nu\rho}-B_{\nu\rho})\xi^\mu\xi^\nu
\right. \nonumber \\
&&\qquad \left. +(D_\mu D_\nu \hat \Phi (\bar X) )\xi^\mu\xi^\nu k
+ (D_\mu D_\nu \hat T (\bar X) )\xi^\mu\xi^\nu \right)\; .
\nonumber 
\end{eqnarray}
All necessary techniques can be found in \cite{braaten85,Osborn91}.

In (\ref{S2}) ${\cal R}$ denotes the target space Riemann tensor 
corresponding 
to the target space metric $G_{\mu\nu} (\bar X)$, $D_\mu$ is the covariant
derivative for that metric, 
\begin{equation}
D_a\xi^\mu =\partial_a\xi^\mu +\gamma_{\nu\rho}^\mu 
(\partial_a \bar X^\rho )\xi^\nu \label{covder}
\end{equation}
with the target space Christoffel connection $\gamma$. The full covariant
derivative
\begin{equation}
\nabla_a\xi^\mu =D_a \xi^\mu +\frac 12 {\epsilon_a}^b
(\partial_b \bar X^\nu ){H^\mu}_{\nu\sigma} \xi^\sigma
\label{nabla}
\end{equation}
also depends on
$H_{\mu\nu\rho}=D_\mu B_{\nu\rho}+D_\nu B_{\rho\mu}+
D_\rho B_{\mu\nu}$. 
With our sign conventions $\epsilon^{N\tau}=1$. 

Requiring vanishing variation of $\xi$ at $\partial 
{\cal M}$ one finds that the evaluation of the one-loop effective 
action should be subject to an appropriate boundary 
condition for $\xi$. From the first term (\ref{S2}), therefore, also 
a contribution with normal derivative $\partial_N = N^a\, 
\partial_a$ will emerge. Adding and subtracting a term, the 
appropriate boundary condition can be written as a symmetric 
operator 
\begin{equation}
{\cal B}\, \xi\, \left|_{\partial\, M}\; = \; 
\left( \nabla_{N}+\frac{1}{2}(\nabla_{\tau }\Gamma 
+\Gamma \nabla_{\tau }) + {\cal S}\right) \xi \right|_{\partial  M}=0
\,, \label{bc2}
\end{equation}
where
\begin{eqnarray}
&&\Gamma_{\mu\nu}=(B-F)_{\mu\nu} \,,\nonumber \\
&&{\cal S}_{\mu\nu}=\frac 14 (\partial_N \bar X^\rho ) \left[
{H^\sigma}_{\rho\mu}(F-B)_{\sigma\nu} +
{H^\sigma}_{\rho\nu}(F-B)_{\sigma\mu} \right] 
\nonumber \\
&&\qquad -\frac 12 (\partial_\tau \bar X^\rho )
\left[ D_{\mu} (F-B)_{\nu\rho} +D_\nu (F-B)_{\mu\rho} \right]
\nonumber \\
&&\qquad -(D_\mu D_\nu \hat \Phi )k -(D_\mu D_\nu \hat T )\; .
\label{bc3}
\end{eqnarray}
It is essential that both the normal and tangential
derivatives in (\ref{bc2}) are fully covariant.
The boundary conditions (\ref{bc2}) have a  similar 
structure as ordinary
Robin (or Neumann) boundary conditions. There 
are, however, some important differences which require a 
detailed discussion.

\subsection{Spectral geometry for boundary conditions with normal and
tangential derivatives}\label{ellipt}

Whereas normal derivative boundary terms have a long 
history, in the mathematical literature the 
study of the spectral geometry for differential operators with
boundary conditions depending on tangential
derivatives was initiated by Gilkey and Smith \cite{GilkeySmith}
\footnote{When the present work was completed 
we have been informed by G.Esposito about an earlier paper
on this subject \cite{Grubb}.}.
For that reason boundary conditions of that type are 
named after these authors \cite{AvEs3}. In the string-related literature
boundary conditions with both normal and tangential derivatives
are usually called generalized Neumann boundary conditions.
Several
first heat kernel coefficients have been calculated for such 
problems  by 
McAvity and Osborn \cite{McO} and by Dowker and Kirsten \cite{DK}.
Avramidi and Esposito \cite{AEearlier,AE,AvEs,AvEs3} lifted some commutativity
assumptions and proved a simple criterion for strong
ellipticity. Also \cite{EV} should be mentioned for a related calculation.

It is usually required that the operator ${\cal D}$ appearing in the
quadratic form of the action should be elliptic. In the present case
this means that both world volume metric and target space metric
should be positive definite. On a compact manifold without 
boundary, 
ellipticity is enough to guarantee certain suitable properties of the
spectrum of ${\cal D}$. For example, the number of non-positive modes 
can be at most finite. In order to ensure
the same properties on manifolds with boundary one should also
assume {\it strong ellipticity}
of the boundary value problem \cite{Gilkey}. Dirichlet and
Neumann (Robin) boundary value problems are
always strongly elliptic. For the more general boundary conditions (\ref{bc2})
strong ellipticity is not fulfilled automatically. One should require
that $|\Gamma|<1$, i.e.\ that  all eigenvalues of $\Gamma$ lie 
between $-i$ and $+i$ \cite{AvEs,AvEs3}. For mathematical
details we refer to the original literature. A simple example is
given in the Appendix B. It is clear from that example that if
strong ellipticity is lost the operator ${\cal D}$ acquires 
 an infinite number of negative modes. 

In the present context the value of $\Gamma$ for which 
strong ellipticity does not hold any more corresponds to critical values
of the gauge field as observed in \cite{Nes}\footnote{Some 
consequences of the existence of such a critical value for 
noncommutative geometry have been considered in 
\cite{SST00,Klebanov00}.}. 
Beyond that point
the semi-classical approximation breaks down. This does not
necessarily mean that no quantum theory exists. Perhaps
the path integration over the negative modes could be performed
non-perturbatively, as it has been done with zero modes 
leading to collective coordinates for excitations around the 
background provided by an  instanton \cite{insttH,instO}.

As long as the boundary value problem is strongly elliptic
($|\Gamma|<1$) the heat kernel expansion is well defined.
There is an asymptotic
series as \( t\rightarrow +0 \) of the form 
\begin{equation}
{\textrm{Tr}}(\exp (-t{\cal D}))=
\sum _{n\geq 0}t^{-m/2+n}a_{n}({\cal D})\,,\label {asymp}
\end{equation}
 where \( n=0,\frac{1}{2},1,\dots  \).   ${\rm Tr}$ denotes the functional
trace in the space of square integrable functions.

For a Laplace type operator
\begin{equation}
{\cal D}=-(\nabla_a\nabla^a +E) \label{defD}
\end{equation}
and the boundary conditions (\ref{bc2}) the heat kernel
coefficient $a_1$ has been calculated in  \cite{McO}:
\begin{eqnarray}
&&a_1 =\frac 1{(4\pi )^{m/2}} \left[
\int_{\cal M}{\rm tr} \left( E +\frac R6 \right) \right. 
\nonumber \\
&&\qquad \left.
+\int_{\partial {\cal M}} {\rm tr} \left( b_0 k +b_2 {\cal S} +\sigma_1 k
\Gamma^2 \right) \right] \,,
\label{a1}
\end{eqnarray}
with
\begin{eqnarray}
 && b_{2}=\frac{2}{1+\Gamma ^{2}}\,,\nonumber \\
 && b_{0}+\sigma _{1}\Gamma ^{2}=\frac{1}{3}\, \,. \label {funcs}
\end{eqnarray}

\subsection{Divergences}

In order to isolate the divergent terms 
 we make use of the \( \zeta  \)-function regularization \cite{zeta}.
The zeta function of an elliptic operator \( D \) is defined as 
\begin{equation}
\zeta _{\cal D}(s)={\textrm{Tr}}({\cal D}^{-s})\, \, .\plabel {defzeta}
\end{equation}
 In term of the zeta function (\ref{defzeta}) the effective action 
reads 
\begin{equation}
W= \lim\limits_{s \to 0}\, \left[ -\frac{1}{2s}\zeta 
_{\cal D}(0)-\frac{1}{2}\zeta'_{\cal D}(0)\, \right]\,,\label {W2}
\end{equation}
 where the prime denotes differentiation with respect to \( s \).

The divergent part of the effective action follows from 
the well-known relation between the zeta function and the heat kernel
coefficients 
\begin{equation}
 \zeta _{\cal D}(0)=a_{1}(1,{\cal D},{\mathcal{B}}) \,. \label{zetaa1}
\end{equation}
In our case $a_1$ is given by (\ref{a1}), (\ref{funcs}).
The endomorphism $E$ can be read off from (\ref{S2}):

\begin{eqnarray}
&&E_{\rho\nu}=\frac 12 \partial_a\bar X^\mu 
\partial_b\bar X^\sigma \epsilon^{ab} D_\rho H_{\nu\mu\sigma}
-\frac 14 \partial_a\bar X^\mu\partial_b\bar X^{\mu'}h^{ab}
{H_{\rho\mu}}^\sigma H_{\sigma\mu'\nu} \nonumber \\
&&\quad +{\cal R}_{\mu\nu\rho\sigma}(\partial_a\bar X^\mu )
(\partial^a\bar X^\sigma ) -\frac 12 R(D_\rho D_\nu \Phi )
-D_\rho D_\nu T \label{Egen}
\end{eqnarray}

Thus the divergent part of the effective action becomes 
\begin{eqnarray}
&&W_{{\mbox {\scriptsize {div}}}}=-\frac 1{8\pi s}
\int_{\cal M} d^2x \sqrt{h} \left[ 
\partial_a \bar X^\mu \partial_b \bar X^\rho h^{ab}
\left( -\frac 14 {H_{\nu\mu}}^\sigma H_{\sigma\rho\nu}
+{\cal R}_{\mu\rho} \right) \right. \nonumber \\
&&\qquad\qquad \left. +\frac 12 
\partial_a \bar X^\mu \partial_b \bar X^\sigma \epsilon^{ab}
D^\nu H_{\nu\mu\sigma} +R\left( \frac d6 -\frac 12
D^\nu D_\nu \Phi \right) -D^\nu D_\nu T \right] \nonumber \\
&&\qquad\quad\ -\frac 1{8\pi s}
\int_{\partial {\cal M}} d\tau \left[ (\partial_N \bar X^\rho )
{H_\sigma}^{\rho\mu} (F-B)^{\sigma\nu} 
\left( 1+(F-B)^2 \right)^{-1}_{\mu\nu} \right. \nonumber \\
&&\qquad\qquad -(\partial_\tau \bar X^\rho )
2D^\mu {(F-B)^\nu}_\rho 
\left( 1+(F-B)^2 \right)^{-1}_{\mu\nu} \nonumber \\
&&\qquad\qquad +k\left( \frac d3 -2 (D^\mu D^\nu \hat \Phi )
\left( 1+(F-B)^2 \right)^{-1}_{\mu\nu} \right) \nonumber \\
&&\qquad\qquad \left. -2 (D^\mu D^\nu \hat T )
\left( 1+(F-B)^2 \right)^{-1}_{\mu\nu} \right] \,,
\label{Wdivold}
\end{eqnarray}
where $d$ is dimension of the target space.

This is the most general expression ever derived, though 
 essential parts of it have been
already given in \cite{dornotto86,Callan88,Osborn91}.
The most remarkable term is the one with  $\partial_N \bar X$:
\begin{equation}
(\partial_N \bar X^\rho )
{H_\sigma}^{\rho\mu} (F-B)^{\sigma\nu} 
\left( 1+(F-B)^2 \right)^{-1}_{\mu\nu} \,, \label{newterm}
\end{equation}
which originates from $b_2S$ in (\ref{a1}).
Such a term is  absent in the bare action (\ref{act}).
This means that the model (\ref{act}) as it stands is at 
least {\it not multiplicatively  renormalizable}. Actually 
that term (\ref{newterm}) had been noticed already before in the one-loop
divergences, as calculated  in \cite{dornotto86,Callan88} and in the
conformal anomaly \cite{Osborn91}.

Clearly one way to eliminate that disturbing term would be 
to assume that $\partial_N \bar X$ can be removed using the 
boundary conditions for the background field \cite{Callan88}. 
This amounted to assuming $\bar X$ to be ``on-shell''.
As argued in Sec. 2, our generalization consists in identifying
$\bar X$ with the {\it mean} field and 
therefore to be arbitrary. Already in \cite{dornotto86} it was suggested 
to introduce a 
new coupling to restore multiplicative renormalizability. 
To deal with boundary divergences it was proposed in \cite{behrndtdorn92}
to give the background fields $G_{\mu\nu},\ B_{\mu\nu}$, etc.
an explicit dependence on the world-volume coordinates.
{}From the sigma model point of view this would correspond to
the introduction of infinitely many new couplings which we
want to avoid.

It would be natural to also consider the question, 
 whether one can add more couplings
with proper dimension and symmetries. The only possibility
seems to be the volume term:
\begin{equation}
\int_{\cal M} d^2 x\sqrt{h} v_\mu^a \partial_a X^\mu \,.
\label{osbcoupl}
\end{equation}
Consequences of this coupling have been studied by Osborn
\cite{Osborn91}. This term would modify the connections
in (\ref{covder}) and (\ref{nabla}) and make the
equation (\ref{Wdivold}) considerably more lengthy,
but would not introduce a qualitative 
change. Another
possible candidate could be a bulk term $K_\mu \partial^2X^\mu$.
However, by integration by parts this term can be absorbed in
a redefinition of $G_{ab}$ and the $\partial_N X^\mu$ coupling.

In quantum field theory any 
``new'' term, appearing as a counter term in the action does not 
necessarily imply nonrenormalizability. E.g.\ a 
regularization scheme which breaks Lorentz invariance would 
introduce such counterterms automatically. Actually the 
situation is not so different here. In principle the 
appearance of many possible types of new counterterms at 
the boundary of a manifold should be possible, because the 
invariances of the theory are broken there.  
We cannot totally exclude the possibility that the counterterm
(\ref{newterm}) can be removed by a field redefinition, but
this seems to be very unlikely. Indeed, the fact the new coupling 
provides a rather beautiful mechanism for 
the emergence of $D$-branes (cf.\ \cite{DLP89,Leigh89} and 
Sec.\ 4.4 below) may represent a strong argument against this 
possibility.
The final 
argument whether a new counter term indeed leads to a 
nonrenormalizable theory must come from its eventual 
effect on some physical observable (S-matrix element with 
on shell external lines\footnote{Of course, if we put the mean
field $\bar X$ on shell, in the conventional sigma model (\ref{act})
the dangerous term $\partial_N\bar X$ in the {\it one-loop}
effective action (\ref{Wdivold})
can be traded for a term containing $\partial_\tau \bar X$
\cite{Callan88}. However, the corresponding divergence can
resurface at higher orders of the loop expansion.}), 
or, if that counterterm produces a 
``genuine'' anomaly relative to some symmetry which is 
broken at the quantum level. 

In our present paper we follow neither of these lines. 
Instead, we regard the appearance of (\ref{newterm}) as an indication 
for the necessity to introduce a new coupling of 
$\partial_NX^\mu$ to another 
vector field $V_\mu$ at the boundary. 
Such a field has been 
proposed before 
\cite{dornotto86,DLP89,Leigh89,behrndtdorn92,dornotto96,Ellis97,Mavromatos99}
though our treatment of that field is different.
At the moment there is
no indication that the new coupling does not affect the physical
content of the theory. On the contrary, it just seems to be 
the missing link between different string/brane models.
In order to investigate its 
consequences in detail we consider its main features in a 
simplified string model in the next Section. 

%%%%%%%%%%%%%%%%%%%%%%%%%%%%%%%%%%%%%%%%%%%%%%%%%%%
\section{Sigma model with $\partial_N X$ coupling}
\subsection{The model}
To simplify the discussion we drop all dimensional
couplings from (\ref{act}) and consider the  action
with the $V$ field present from the start
\begin{eqnarray}
&&S= \int _{\cal M}d^{2}z\sqrt{h}
\left(\frac{1}{2} G_{\mu\nu} h^{ab}
\partial _{a}X^{\mu }\partial _{b}X^{\nu }
+\frac{1}{2\sqrt{h}}\epsilon^{ab}B_{\mu\nu}
\partial _{a}X^{\mu }\partial _{b}X^{\nu } \right)
\nonumber \\
&&\qquad+\int _{\partial {\cal M}}d\tau \left(
 A_{\mu }(X)\partial _{\tau }X^\mu
 -V_{\mu}(X)\partial _{N}X^\mu
\right) \, .
\label{actnew}
\end{eqnarray}

The invariances of the new term can be most easily seen if 
it is represented as a bulk integral
\begin{equation}
\int_{\cal M} d^2 x\sqrt{h} \nabla_R^a V_\mu\partial_a 
X^\mu\; ,
\label{bt}
\end{equation}
where $\nabla_R^a$ is the Riemannian covariant derivative on
${\cal M}$. Since $X^\mu$ is a scalar on the world sheet, the
expression under the integral (\ref{bt}) is a scalar density.
Variation of of the action (\ref{bt}) under the diffeomorphism
$x^a\to x^a +\eta^a$ is therefore given by the surface integral:
\begin{equation}
\int_{\partial {\cal M}} d\tau \eta^N 
\nabla_R^a V_\mu\partial_a X^\mu \,.\label{varbt}
\end{equation}
This variation vanishes for $\eta^N\oB =0$, i.e. for the
diffeomorphisms which do not shift the boundary. In particular,
the new coupling is invariant under the one-dimensional
diffeomorphism transformations of the boundary, in agreement 
with the standard property of diffeomorphism invariant actions. 

There is no explicit gradient symmetry associated with $V_\mu$.
However, the introduction of a transformation similar to 
a gauge transformation could be envisaged, 
involving $V_\mu$ in the same way as it has been done with
the coupling (\ref{osbcoupl}) (cf.\ eq.\ (4.3) of \cite{Osborn91}).
Since any symmetry of such a type will be of no consequences 
for our present work we will not persue this point any 
further here. On the other hand, an argument against the interpretation of 
$V_\mu$ as a gauge field could be that, from the point of 
string theory, it prepresents an off-shell degree of freedom 
whose physical content is still to be determined.

Proceeding as before for the general model we arrive at

\begin{eqnarray}
&&S_2 =-\frac 12 \int _{\cal M}d^{2}z\sqrt{h} \xi^\mu (\nabla^a
\nabla_a +E)_{\mu\nu} \xi_\nu +\frac 12 \int _{\partial {\cal M}}d\tau 
\left(-\xi^\nu (G_{\nu\mu}+2D_\nu V_\mu )\nabla_N \xi^\mu  \right.
\nonumber \\
&&\quad -(D_\rho D_\sigma V_\mu )(\partial_N\bar X^\mu )\xi^\rho
\xi^\sigma +2(\partial_\tau \bar X^\lambda )(D_\nu V_\mu )
{H^\mu}_{\lambda\sigma}\xi^\nu \xi^\sigma \label{S2new} \\
&&\quad  \left. +\xi^\nu D_\tau \xi^\mu (B_{\mu\nu}-F_{\mu\nu}) +
(\partial_\tau \bar X^\rho ) D_\mu (F_{\nu\rho}-B_{\nu\rho})\xi^\mu\xi^\nu
\right) \nonumber
\end{eqnarray}
where $E$ coincides with (\ref{Egen}) when $\Phi =T=0$.
The covariant derivative $\nabla$ is given in (\ref{nabla}).

The natural inner product in the space of fluctuations reads:
\begin{equation}
<\xi_{(1)},\xi_{(2)}>=\int_{\cal M}d^2 x\sqrt{h} G_{\mu\nu} (\bar X)
\xi_{(1)}^\mu \xi_{(2)}^\nu
\label{innprod}
\end{equation}

\subsection{Boundary conditions}
\label{bcsection}
To discuss the boundary conditions we now need a somewhat 
more general setting. 
The main features of (\ref{actnew}) are reflected in the structure  
\begin{equation}
S=\frac 12 \int_{\cal M} ((\nabla \xi )^2 -\xi E \xi )
+\frac 12 \int_{\partial {\cal M}} 
(-\xi \Lambda \nabla_N \xi +\xi L \xi )\,,
\label{S2gen}
\end{equation}
of the action, where $E$ and $\Lambda$ are the appropriate matrix-valued 
functions on ${\cal M}$ and $\partial {\cal M}$, respectively, and
$L$ is a differential operator on the boundary $\partial {\cal M}$.
We split $L$ into a hermitian and an antihermitian piece:  

\begin{equation}
L=L_++L_-\,,\qquad L_\pm =\pm L_\pm^\dag \,. \label{split}
\end{equation}
Note, that $L_+$ is fixed by the action (\ref{S2new}) while
$L_-$ is arbitrary, since it drops out from 
(\ref{actnew}) or (\ref{S2gen}).  $E$ is supposed to be hermitian.

At first we consider  $\Lambda =0$. In this case there is a convenient way to
derive the spectral problem corresponding to the action
(\ref{S2new}). Variation of the action (\ref{S2new}) with
respect to $\xi$ leads to 
\begin{equation}
\delta S=- \int_{\cal M}  (\delta\xi )(\nabla^2 +E)\xi
+\int_{\partial {\cal M}} (\delta\xi )
(-\nabla_N \xi +L_+\xi ) \,.\label{varS2new}
\end{equation}
The bulk integral in (\ref{varS2new}) contains the relevant
elliptic operator ${\cal D}=-(\nabla^2 +E)$, while the surface
part defines the boundary conditions which can be either
of Dirichlet type

\begin{equation}
\xi \oB =0 \label{Dir}
\end{equation}
(following from $\delta\xi \vert_{\partial M}  =0$ after absorption
of an inhomogeneous mode of $\xi$ into the background),
or of modified Neumann type
\begin{equation}
(-\nabla_N \xi +L_+\xi ) \vert_{\partial M} =0 \label{Neu} \,.
\end{equation}
For both cases (\ref{Dir}) and (\ref{Neu})
two  important properties are to be noted:  (i) the operator
${\cal D}$ is formally self-adjoint with respect to (\ref{innprod}):
\begin{equation}
<\xi_{(1)},{\cal D}\xi_{(2)}>=<{\cal D}\xi_{(1)}, \xi_{(2)}>\, ,
\label{self}
\end{equation}
and (ii) after integration by parts in (\ref{S2new}) 
or (\ref{S2gen}) the surface
term vanishes. Due to (i) the operator ${\cal D}$ possesses a complete set of
orthonormal eigenfunctions $\psi_n$ corresponding to the eigenvalues
$\lambda_n$: $\psi (x) =\sum_n \psi_n(x)c_n$. 
Due to (ii) the action (\ref{S2new}) takes the form
$S=\frac 12 \sum_n c_n^2 \lambda_n$, and the path integral 
for the effective action $W$ becomes 
\begin{equation}
W=-\ln \left( \prod_n \lambda_n \right)^{-1/2}=
-\ln \left( \det {\cal D} \right)^{-1/2} \label{pathint}
\end{equation}  
where we assume a suitable regularization of the infinite
product. 

If $\Lambda \ne 0$ the procedure described above is not as 
easy to apply. The reason is that on the boundary
$\delta\xi$ and $\delta\nabla_N\xi$ are independent.
However, we can still select 
two relevant sets of the boundary conditions by requiring 
(i) and (ii) to hold.
It is clear that (ii) is true for Dirichlet boundary
conditions (\ref{Dir}) and for modified Neumann
conditions
\begin{equation}
(-(1+\Lambda )\nabla_N\xi +L\xi )\vert_{\partial M} =0 \,.\label{Neu2}
\end{equation}
To satisfy (i) we should require
\begin{equation}
((1+\Lambda )^{-1}L)^\dag =(1+\Lambda )^{-1}L\,, \label{herm}
\end{equation}
i.e. the operator $(1+\Lambda )^{-1}L$ should be hermitian
with respect to the restriction of (\ref{innprod})
to the boundary. The equation (\ref{herm}) will be used
to define $L_-$ in the decomposition
(\ref{split}).

{}From the mathematical point of view both conditions
(\ref{Dir}) and (\ref{Neu2}) are admissible and lead
to a consistent quantum field theory. From the physical
point of view the conditions (\ref{Neu2}) are much
superior. The first reason is that it does not seem natural
for the boundary conditions to be completely independent
of the boundary couplings. The second and  more
important one is that the  Dirichlet conditions
(\ref{Dir}) are in fact contained as a special case in the 
generalized Neumann ones 
(\ref{Neu2}). This property of the boundary conditions
(\ref{Neu2}) (which provides a mechanism of spontaneous creation
of $D$-branes) will be discussed in sec. \ref{Dbranes}.

We conclude this section by comparing our boundary conditions
with the ones in the existing literature. 
For duality reasons the coupling $V_\mu\partial_N X^\mu$
has been introduced in \cite{DLP89,Leigh89}\footnote{
A coupling to a normal derivative at the boundary has been
introduced also in order to include recoil operators of the D-brane
in a description of scattering, following the Liouville approach 
\cite{Ellis97,Mavromatos99,Ellis98}.}. However,
a different set of mixed boundary conditions has been
proposed \cite{Leigh89}
\begin{eqnarray}
&&f^\mu_{,A}G_{\mu\nu}\partial_NX^\mu +(B-F)_{AB}\partial_\tau \eta^B 
\oB =0 \nonumber \\
&&V_\mu \oB =0 \label{leighbc}
\end{eqnarray}
instead of our conditions (\ref{Neu2}). Here $\eta^A$ are coordinates
on the $D$-brane, $f^\mu_{,A}$ are projectors on the $D$-brane
directions. The boundary conditions (\ref{leighbc}) are fully
legitimate and lead to a self-consistent sigma model. Of course,
to obtain (\ref{leighbc}) the $D$-brane need be introduced 
from outside.
In our approach it appears spontaneously in the open string sigma
model (see Sec.\ \ref{Dbranes}). Also, as it is clear from 
Sec. \ref{effective}, if one starts with the conventional
open string sigma model ($V=0$) and then turns on the $V$-field
by making insertions in the propagator, one would arrive at
our boundary conditions (\ref{Neu2}) without any further assumptions.

\subsection{Beta functions}
To calculate the beta functions the
operator $L_-$ in the decomposition (\ref{split}),
as defined by the condition (\ref{herm}), must be determined first. Let
$(1+\Lambda )^{-1}=1+l$. Then (\ref{herm}) yields
\begin{equation}
L_-=\frac 12 \left( L_+l^\dag -lL_+ -L_-l^\dag -lL_-
\right) \,, \label{herm2}
\end{equation}
which  admits an iterative solution.
Let $L^{(n)}_-$ be a part of $L_-$ of $n$th order in $l$
so that $L_-=\sum_{n=0}^\infty L^{(n)}_-$. With
\begin{eqnarray}
&&L_-^{(0)}=0 \\
&&L_-^{(1)} =\frac 12 \left( L_+l^\dag -lL_+ \right)
\label{L1}\end{eqnarray}
one gets the $n$-th term
\begin{equation}
L_-^{(n)} =-\frac 12 \left( L_-^{(n-1)}l^\dag +
lL_-^{(n-1)} \right)\, ,
\end{equation}
and thus
all finite order contributions $L_-^{(n)}$ are easily determined
in powers of $l$ and $l^\dag$,
for example,
\begin{eqnarray}
&&L_-^{(2)} =-L_+ \left( \frac{l^\dag }2 \right)^2
+\left( \frac{l }2 \right)^2 L_+ \,, \\
&&L_-^{(3)} =L_+ \left( \frac{l^\dag }2 \right)^3
+\left( \frac{l }2 \right) L_+ \left( \frac{l^\dag }2 \right)^2
-\left( \frac{l }2 \right)^2 L_+ \left( \frac{l^\dag }2 \right)
-\left( \frac{l }2 \right)^3 L_+ \; .
\end{eqnarray}
It is easily verified that the integral
\begin{equation}
L_-=\frac 12 \int_0^\infty da\, e^{-a} \exp \left( -\frac{la}2 \right)
\left( L_+l^\dag -lL_+ \right) \exp \left( -\frac{l^\dag a}2 \right)
\label{compL-}
\end{equation}
summarizes $L_-$ to all orders in $l$ and $l^\dag$.

The strategy for the calculation of the heat kernel coefficient
$a_1$ is rather simple. Given $L_-$ one should re-write
the boundary condition as
\begin{equation}
(-\nabla_N +(1+\Lambda )^{-1} L )\vert_{\partial M}  =0 \label{bcLam}
\end{equation}
and then transform it to the form (\ref{bc2}) thus defining the matrices 
$\Gamma$ and $S$. $a_1$ then follows from (\ref{a1}). 

When $l=(1+\Lambda )^{-1}-1$ is expanded in terms of $\Lambda$,
the first order contribution is  $l_{\mu\nu}=
-2(D_\mu V_\nu )$. Since $V$ is physically a more relevant
expansion parameter than $l$, we re-expand all terms in $a_1$ (\ref{a1})
to first order of $V$. After long but otherwise elementary
calculations from (\ref{S2new}), (\ref{S2gen}),
(\ref{split}), (\ref{L1}), (\ref{W2}), (\ref{zetaa1}) and
(\ref{a1}) the divergent part of the effective action at $\partial{\cal M}$
is obtained:
\begin{eqnarray}
&&W_{{\mbox {\scriptsize {div}}}}[\partial{\cal M}]=-\frac 1{4\pi s}
\int_{\partial{\cal M}} d\tau (1+\Gamma_0^2)_{\mu\nu}^{-1}
\left[ (\partial_\tau \bar X^\rho ) \left\{
S^{(\tau )}_{\rho ,\nu\mu} -2 (D_\nu V_\sigma ){H^\sigma}_{\rho\mu}
\right.\right. \nonumber \\
&& \qquad\qquad
-2 (\Gamma_{0,\nu\sigma} (D_\sigma V_\lambda ) 
\Gamma_{0,\lambda\alpha} + (D_\nu V_\sigma )
\Gamma^2_{0,\sigma\alpha}) (1+\Gamma_0^2)^{-1}_{\alpha\beta}
S^{(\tau)}_{\rho ,\beta \mu} \nonumber \\
&& \qquad\qquad \left. +(D_\rho D_\nu V_\sigma )\Gamma_{0,\sigma\mu}
-2 (D_\nu V_\sigma )S^{(\tau )}_{\rho ,\sigma\mu} \right\}
\nonumber \\
&& \qquad\qquad +(\partial_N \bar X^\rho )\left\{ S^{(N )}_{\rho ,\nu\mu}
-2 (D_\nu V_\sigma )S^{(N )}_{\rho ,\sigma\mu} 
+(D_\nu D_\mu V_\rho ) 
\right. \nonumber \\
&& \qquad\qquad \left.\left.
-2 (\Gamma_{0,\nu\sigma} (D_\sigma V_\lambda ) 
\Gamma_{0,\lambda\alpha} + (D_\nu V_\sigma )
\Gamma^2_{0,\sigma\alpha}) (1+\Gamma_0^2)^{-1}_{\alpha\beta}
S^{(N)}_{\rho ,\beta \mu} \right\} \right] \,. \label{Wdivnew}
\end{eqnarray}
The bulk  terms are the same as
in (\ref{Wdivold}). The abbreviations
\begin{eqnarray}
&&S^{(N)}_{\rho ,\mu\nu} =\frac 14 [ {H^\sigma}_{\rho\mu}
(F-B)_{\sigma\nu} + \ \mu \leftrightarrow \nu  ] \,,
\nonumber \\
&&S^{(\tau )}_{\rho ,\mu\nu}=-\frac 12 [ D_\mu (F-B)_{\nu\rho}
+ \ \mu \leftrightarrow \nu  ] \,,\nonumber \\
&&\Gamma_{0,\mu\nu}= (B-F)_{\mu\nu} \,. \label{nots}
\end{eqnarray}
have been introduced.
The beta functions $\beta_\rho^A$ and $\beta_\rho^V$ 
to linear order in $V$ can be
easily extracted from (\ref{Wdivnew}). It should be recalled that
from all background couplings we had retained $A$, $B$ and $V$
only. The target space geometry is assumed to be flat.
Therefore no distinction between upper and lower
indices in (\ref{Wdivnew}) need to be made and the
covariant derivative $D_\mu$ (\ref{covder}) coincides with
the partial derivative $\partial_\mu$.

There is one particular case which can be analysed without
additional calculations. It corresponds to constant $B$-field or
$H_{\mu\nu\rho}=0$.
It is clear that in this case the $0$th order (in $l$) 
contribution to the beta functions are given by
the expressions with $V=0$ derived in the previous
section. To this order  
$\beta_\mu^V$ vanishes if $H_{\mu\nu\rho}=0$ . 
Therefore,
the equation $\beta_\mu^V=0$ has a solution $V=0$.
This can be checked with the help of equation (\ref{Wdivnew}).
Then the beta function $\beta_\mu^A$ is given by its
old expression and is proportional to variation
of the Born-Infeld action with respect to $A$.
Hence, for  constant $B$ field we reproduce the
Born-Infeld action as one of the possible solutions.
There could be also non-perturbative ones when $V$ is
not small. Also, at least for small $V$, the expansion
in $V$ or $l$ corresponds to an expansion in derivatives of
the $B$ field.

%%%%%%%%%%%%%%%%%%%%%%%%%%%%%%%%
\subsection{Spontaneous creation of $D$-branes}\label{Dbranes}
The boundary
conditions (\ref{Neu2})
\begin{equation}
\left[ (\delta_{\mu\nu}+2D_\mu V_\nu )\nabla_N
 +L_{\mu\nu} \right]\xi^\nu \oB =0 \label{finVbc}
\end{equation}
have intriguing qualitative features. Let us consider the simplest case when
$D_\mu V_\nu =D_\nu V_\mu$, $[DV,\Gamma_0 ]=0$, where $\Gamma_0=B-F$.
If we rewrite  (\ref{finVbc}) in the form
(\ref{bcLam}) the operator $\Gamma$ becomes
\begin{equation}
\Gamma = (1+2DV )^{-1}\Gamma_0 \,. \label{Gam}
\end{equation}
When $(1+2DV)^{-1}$ is sufficiently large
so that the absolute value of some of the eigenvalues
of $\Gamma$ becomes larger than $1$, strong ellipticity
is lost, and, as discussed in sec. \ref{ellipt}, we
enter a non-perturbative strong coupling regime.
There an infinite number of negative modes
appears which correspond to tachyonic modes in
the Minkowski signature theory.

On the other hand,
when the matrix $(1+2DV )$ becomes degenerate, so that
$(1+2DV )\xi^{[0]}=0$ for some 
$\xi^{[0]}\ne 0$,  (\ref{finVbc}) implies
\begin{equation}
L\xi^{[0]}\oB =0 \label{Dir21}
\end{equation}
or, since for generic values of the other background couplings
$A$, $B$, etc., 
$L$ is non-degenerate --this should be true at least
on the zero-measure subset where $(1+2DV )$ is degenerate--
\begin{equation}
\xi^{[0]}\oB =0 \label{Dir31}\,.
\end{equation}
This clearly indicates the formation of a $D$-brane.
The co-dimensionality of that $D$-brane is equal to number
of zero eigenvalues of $(1+2DV)$. Note  that
at this point strong ellipticity is recovered again,
but with a different type of boundary conditions.
We conclude that the model with $V$-coupling
describes both open strings and $D$-branes depending
on the value of the derivatives of the field $V$.

Multiple roots of $(1+2DV)$ correspond to multiple
$D$-branes sitting on top of each other. Such configurations
should produce non-abelian Chan-Paton factors 
\cite{Polchinski95}.

With some restrictions on the form of the background
couplings it is possible to calculate the $\beta$-functions
without expanding in power series of $V$. In this way
a clearer picture of  $D$-brane formation can be obtained.
We postpone this task for a future publication.

The present mechanism of $D$-brane formation should be compared
with the one proposed and studied in the previous literature
\cite{Sen99,Harvey00,HKM00,Majumder00,Mello00,Moeller00,Husain00}.
The mechanism proposed there  in our framework  corresponds
to a dominating $L$ in the boundary conditions
(\ref{finVbc}) (with $V=0$), so that this condition
(\ref{finVbc}) becomes approximately Dirichlet and an exact Dirichlet 
boundary condition could be achieved for infinite $L$
only. In our case Dirichlet boundary conditions appear
for finite background couplings due to the vanishing of the
coefficient in front of the normal derivative. Of course,
in the framework of our approach  also 
a combination of the two mechanisms could be envisaged.

%%%%%%%%%%%%%%%%%%%%%%%%%%%%%%%%%%%%%%%%%%%%
\section{Conclusions}
We have devoted much space (Sec. 2 and Appendix A) to the
consideration of the effective action in the presence
of boundaries. It seems that a completely satisfactory
construction of such an object is still missing in the
literature.  We do not pretend to have obtained  the ultimate
solution as well. However, we were able to  extract two important lessons
from our considerations. First,
we have seen that to be able to reproduce reliably
all divergences of the 1-p-i diagrams one should not
impose any restriction on the mean field. In particular,
the latter should not satisfy any boundary conditions.
Second, we have shown how the boundary conditions are
changed due to ``perturbative'' insertions of boundary
interactions in the propagator. In rederiving
the divergent part of the effective action
for conventional open string sigma model some mathematical
issues like strong ellipticity of the boundary value problem
have been clarified. Also, we were able to obtain exact results
for slightly more general couplings than the ones exhibited
in previous calculations.
We also reconsidered the approach of \cite{Callan88} where
the mean field $\bar X$ was taken on-shell. While this
worked perfectly for the renormalization of the
one-loop physical amplitudes, to analyse multiplicative
renormalizability of the full theory we suggest to
to extend the method of \cite{Callan88}
 to an off-shell mean field $\bar X$ and so that
classical boundary conditions are no longer relevant for it. 
Our analysis strongly suggests the
introduction of a new coupling $V_\mu \partial_N X^\mu$
at the boundary. In fact, such  a step has been proposed
long ago \cite{dornotto86}. 

Therefore, the sigma model with that new
coupling has been studied. After formulating the boundary
value problem,   the $\beta$-functions to the
leading order in $V$ were calculated. 
For constant $B$-field we found that one of the beta functions,
$\beta^V$, vanishes for $V=0$ while the other, $\beta^A$, then reproduces
the variation of the Born-Infeld action. The perturbative expansion
in terms of
small $V$ generates derivative corrections to that action\footnote
{Derivative corrections to the Born-Infeld action for vanishing or constant
$B$-field as power series in the $F_{\mu\nu}$ generated by higher
loops were calculated in \cite{AT1,AT2}. Recently the emergence of such
corrections in non-commutative geometry was also considered
\cite{Okawa00,Cornalba99}.}. 
It would be interesting to compare the derivative
corrections obtained by the beta function method with the ones following
from the Seiberg-Witten transformation \cite{Seiberg99}. 
Of course, this would require a considerable improvement of our
understanding of non-commutative geometry in the presence of a
non-constant $B$-field (see \cite{Ho00} for some recent results
and further references).

To us the most exciting feature of the new coupling is that a natural 
mechanism
of spontaneous $D$-brane creation
is implied. As the matrix $(\delta_{\mu\nu}+
2D_\mu V_\nu )$ becomes degenerate the boundary conditions (\ref{finVbc})
are transformed to  mixed Dirichlet-Neumann ones,
thus describing a $D$-brane with co-dimensionality equal to the
number of zero eigenvalues of that matrix. 

In our view, the $V$-coupling deserves further investigation.
In particular,
one should clarify the duality properties of the modified
sigma model. Even though our model (\ref{actnew}) for the string with
boundary terms looks very similar
to the one proposed in \cite{DLP89,Leigh89}, the boundary conditions
are different, so that $T$-duality transformations may require
modification. Of course, another interesting generalization of our
results would be the inclusion of supersymmetry.  

%%%%%%%%%%%%%%%%%%%%%%%%%%%%%%%%%%%%%%%%%%%%
\section*{Acknowledgements}

This work has been supported by the FWF (Fonds
zur F\"{o}rderung der wissenschaftlichen Forschung) 
project P-12.815-TPH,
the Alexander von Humboldt Foundation and the DFG, grant Bo 1112/11-1.
One of the authors (DV) is grateful to the Erwin Schr\"{o}dinger
Insitute for Mathematical Physics for warm hospitality.
We are grateful to  A. Andrianov,
I.Avramidi, I.Bandos, G.Esposito, M.Kreuzer and H.Dorn
for discussions and thank the referee for constructive remarks. 

%%%%%%%%%%%%%%%%%%%%%%%%%%%%%%%%%%%%%%%%%%%%%

\section*{Appendix A:
One-loop effective action and classical background field
(a formal argument)}

1-p-i-vertices 
may be extracted from the effective action $W\, 
(\bar{\phi}_i)$ where $\bar\phi$ is the ``mean field'', 

\begin{equation}
\bar{\phi_i} \; = \; \frac{\delta\,Z_c}{\delta\, j_i}\; ,
\label{w1} 
\end{equation}
related to the generating functional $Z\, (j)$ of 
Green-functions $Z = \exp i Z_c$ by differentiation with 
respect to sources $j_i$ in the latter. We use deWitt's 
notation with the index $i$ describing the space-time 
variable as well as (eventual) indices of internal 
symmetries and in this Appendix we work in Minkowski space. 
Actually the action $S = S^{{\cal M}} + S^{\partial {\cal 
M}}$ in $Z$ may also contain a source term at the boundary. 

The relation of $\bar{\phi}_i$ to a generic classical 
solution $\ul{\phi}_i$, around which quantum fluctuations 
$\varphi_i$ may 
occur, is not immediate, especially when the quantum system ---  and hence 
$\ul{\phi}_i$ --- obeys certain  boundary conditions

\begin{equation}
F_{\alpha}\,(\phi)\; = \; 0 \label{w2}
\end{equation}
which must be imposed. 
A dependence on external fields is not indicated 
explicitly. 
For our present purposes it is sufficient to consider 
theories without gauge invariance, where $Z\,(j)$ may be 
simply generalized as

\begin{equation}
Z (j,k) = \int (d \phi) (d \lambda)\, C\; e^{ i (\frac{S}{\hbar} + 
j_i\, \phi_i + \lambda_\alpha\, F_{\alpha}\,(\phi) + 
\lambda_\alpha k_\alpha)} \label{w3}
\end{equation}
to include (\ref{w2}) by means of a Lagrange multiplier field 
$\lambda_\alpha$ (Greek indices indicate that their 
space-time variable and related integrals refer to the 
boundary $\partial\,{\cal M}$). It is then convenient to introduce 
a source $k_\alpha$ for that field too (thus restricted to 
$\partial \, {\cal M}$ as well)\footnote{In order to be able to
reproduce the Green functions for fields at $\partial{\cal M}$,
$j_i$ may contain $\delta$-function contributions.}. 
The factor  $C$ should be  
determined by the requirement that the actual path integral 
should involve only the modes obeying the boundary 
conditions. Its derivation in the fully general case 
would imply the development of a Hamiltonian formalism, 
including boundary constraints, adapted to the present 
situation which does not seem to be available. Certain 
similarities to the Faddeev-Popov determinant can be 
expected. A formulation like (\ref{w3}) has the 
advantage that the boundary conditions  are completely 
integrated into the formalism and not imposed ``from the 
outside''. 

The expansion $\phi_i = \ul{\phi_i} + \varphi_i$ around a 
``background'' solution $\ul{\phi_i}$ in (3) requires the 
``saddle point'' condition for the linear term in 
$\varphi_i$ and $\lambda_\alpha$
\begin{eqnarray}
\frac{1}{\hbar}\; \frac{\delta\, S}{\delta\, 
\ul{\phi_i}} \; + \; j_i \; & = & \, 0 \; ,\label{w4} \\
F_{\alpha}\, (\ul{\phi}) \; + \; k_\alpha \; & = & \; 0 \label{w5}
\end{eqnarray}
which {\em only for  vanishing sources} $j_i, \, k_\alpha$ implies that 
the solutions $\ul{\phi_i}$ are the ones of the classical 
e.o.m.-s, subjected to the boundary conditions (\ref{w2}). 
Retaining in (\ref{w3}) only quadratic terms in 
$\varphi_A = (\varphi_i, \lambda_\alpha)$  after some 
simple formal manipulations yields for the mean field (\ref{w1}) 
and an analogous one related to $k_\alpha$ 

\begin{eqnarray}
{\bar\phi}_i \; & = & \; \ul{\phi_i} \, (j,k) \; + \; {\cal 
O}\, (\hbar) \; ,\label{w14} \\
\bar{\lambda}_\alpha \; & = & \; {\cal O}\,(\hbar) \; .
\label{w15}
\end{eqnarray}

Thus to leading order, (one loop in the effective action) 
the mean field is identical to $\ul{\phi_i}\, (j,k)$, but as 
arbitrary as the dependence of the latter on the sources 
$j$ and $k$. Therefore, if the evaluation of the path 
integral in $\varphi$ around $\ul{\phi}$  is performed, the 
simple replacement (\ref{w14}) yields the contribution to 
the effective action. 

In contrast to the argument of Section 2, the result 
(\ref{w14}) is found to hold for a generic action $S$. 
However, it should be noted that the presence of a finite 
boundary $\partial\, {\cal M}$ really forbids the 
application of the usual rules of functional derivation, 
e.g.\ in the expansion of (\ref{w2}). Naive application in 
the case of a normal derivative contained in the boundary 
would lead to a surface term which rides on top of the 
boundary again and thus produces $\delta (0)$. Therefore,  
such terms must be dropped explicitly. Whether this can be 
made into a general consistent prescription must be the 
subject of a more profound analysis.

%%%%%%%%%%%%%%%%%%%%%%%%%%%%%%%%%%%%%
\section*{Appendix B: Strong ellipticity (an example)}
For the Laplace operator
\begin{equation}
\Delta =-\partial_1^2 -\partial_2^2 
\label{Lap}
\end{equation}
on a flat manifold ${\cal M}=[0,1]\times S^1$  
with the boundary conditions
\begin{equation}
\partial_1 \phi |_{x^1=0}=0 \,,\qquad
(\partial_1 +i\alpha \partial_2 )\phi |_{x^1=1}=0 
\label{bctoy}
\end{equation}
there are two sets of eigenmodes. 
The first one is:
\begin{equation}
\phi_{k_1k_2}=\exp (ik_2x^2) \cos (k_1x^1) 
\label{set1}
\end{equation}
These modes satisfy (\ref{bctoy})
at $x^1=0$. The condition at $x^1=1$ should be used
to define discrete values of $k_1$. $k_2$ is quantized
to assure periodicity in the $x^2$ coordinate. 
Obviously, the eigenvalues $k_1^2+k_2^2$
of $\Delta$ are  always non-negative.
The second set of the eigenfunctions 
\begin{equation}
\bar \phi_{k_1k_2}=\exp (ik_2x^2) \cosh (k_1x^1) 
\label{set2}
\end{equation}
again satisfies  (\ref{bctoy})
at $x^1=0$. At $x^1=1$ that condition  yields
\begin{equation}
k_1 \tanh (k_1)=\alpha k_2 \,. 
\label{bcat1}
\end{equation} 
For positive $\alpha k_2$ (\ref{bcat1}) always has two 
solutions. For sufficiently large $|k_2|$ they are
\begin{equation}
k_1 \approx \pm \alpha k_2 
\label{solbc}
\end{equation}
and the corrections are exponentially small. From 
(\ref{solbc}) the eigenvalues $-k_1^2+k_2^2$
of  $\Delta$ are now positive
if $|\alpha|< 1$ and negative if $|\alpha| > 1$. We conclude
that if strong ellipticity of the boundary value problem
is lost ($|\alpha| > 1$) the operator $\Delta$ shows infinitely
many negative modes.

%%%%%%%%%%%%%%%%%%%%%%%%%%%%%%%%%%%%%


\begin{thebibliography}{99}

\bibitem{FT85}E.S. Fradkin and A.A. Tseytlin, 
Phys. Lett. \textbf{163 B} (1985) 123. 
\bibitem{T86} A.A. Tseytlin, Nucl. Phys. \textbf{B 276} (1986) 391.

\bibitem{ACNY87} A. Abouelsaood, C.G. Callan, C.R. Nappi and S.A. Yost, 
Nucl. Phys. \textbf{B
280 {[}FS 18{]}} (1987) 599.

\bibitem{Callan88}
C.G. Callan, C. Lovelace, C.R. Nappi and S.A. Yost,
Nucl. Phys. {\bf B 308} (1988) 221.

\bibitem{CFMP85}C.G. Callan, D. Friedan, E.J. Martinec and M.J. Perry, 
Nucl. Phys. \textbf{B 262} (1985) 593. 

\bibitem{T99}A.A. Tseytlin, Born-Infeld action, 
supersymmetry and string theory, hep-th/9908105,
to appear in the Yuri Golfand memorial volume, ed. M. Shifman, 
(World Scientific,
2000).
\bibitem{dornotto86}
H. Dorn and H.-J. Otto, Z. Phys. {\bf C32} (1986) 599.
\bibitem{DLP89}
J. Dai, R.G. Leigh, J. Polchinski, Mod. Phys. Lett. {\bf A 4}
(1989) 2073.
\bibitem{Leigh89}
R.G. Leigh, Mod. Phys. Lett. {\bf A 4} (1989) 2767.
\bibitem{behrndtdorn92}
K. Behrndt and H. Dorn, Int. J. Mod. Phys. {\bf A7}
(1992) 1375.
\bibitem{dornotto96}
H. Dorn and H.-J. Otto, Phys. Lett. {\bf B 341} (1996) 81.

\bibitem{Gilkey} P.B. Gilkey, 
\textit{Invariance Theory, the Heat Equation, and the Atiyah-Singer
Index Theorem}, (CRC Press, Boca Raton, 1994).

\bibitem{Osborn91}
H. Osborn, Nucl. Phys. {\bf B 363} (1991) 486.

\bibitem{McO2}
D.M. McAvity and H. Osborn, Nucl. Phys. {\bf B 394} (1993) 728.
\bibitem{Symanzik}
K. Symanzik, Nucl. Phys. {\bf B 190} [FS3] (1981) 1.

\bibitem{Wipf94}
C. Wiesendanger and A. Wipf, Ann. Phys. {\bf 233} (1994)
125.

\bibitem{Howe88} P.S. Howe, G. Papadopoulos and K.S. Stelle,
Nucl. Phys. {\bf B 296} (1988) 26.


\bibitem{AB} 
A.A. Andrianov and L. Bonora, 
Nucl. Phys. \textbf{B 233} (1984) 232; 247.
\bibitem{braaten85}
E. Braaten, T.L. Curtright and C.K. Zachos,
Nucl. Phys. {\bf B 260} (1985) 630.


\bibitem{GilkeySmith} P.B. Gilkey and L. Smith, 
The twisted index problem for manifolds with boundary,
J. Diff. Geom. 
\textbf{15} (1983) 393.
\bibitem{Grubb}
G. Grubb, Annali della Scuola Normale Superiore di Pisa
{\bf 1} (1974) 1.

\bibitem{AvEs3} 
I. Avramidi and G. Esposito,
Trends in Mathematical Physics: Proceedings. 
Edited by V. Alexiades and G. Siopsis.
Cambridge, Mass., International Press, 1999. 
(AMS/IP Studies in Advanced Mathematics,
v. 13). pp. 15-34. 
\bibitem{McO} D.M. McAvity and H. Osborn, Class. Quantum Grav. 
\textbf{8} (1991) 1445. 
\bibitem{DK} J.S. Dowker and K. Kirsten, Class. Quantum Grav. 
\textbf{14} (1997) L169; \textbf{16}
(1999) 1917.
\bibitem{AEearlier} I. Avramidi and G. Esposito, Class. Quantum Grav. 
\textbf{15} (1998) 1141. 
 
\bibitem{AE} I. Avramidi and G. Esposito, Class. Quantum Grav. 
\textbf{15} (1998) 281.  
\bibitem{AvEs} I.G. Avramidi and G. Esposito, 
Commun. Math. Phys. \textbf{200} (1999) 495.
\bibitem{EV} E. Elizalde and D.V. Vassilevich, 
Class. Quantum Grav. \textbf{16} (1999) 813.

\bibitem{Nes}
V.V. Nesterenko, Int. J. Mod. Phys. \textbf{A 10} (1989) 2627.
\bibitem{SST00}
N. Seiberg, L. Susskind and N. Toumbas,
{\it Strings in Background Electric Field, Space/Time
Noncommutativity and a New Noncritical String Theory},
hep-th/0005040.
\bibitem{Klebanov00}
I. Klebanov and J. Maldacena, {\it 1+1 dimensional NCOS and
its $U(N)$ gauge theory dual}, hep-th/0006085.

\bibitem{insttH}
G. 't Hooft, Phys. Rev. {\bf D 14} (1976) 3432.
\bibitem{instO}
H. Osborn, Ann. Phys. {\bf 135} (1981) 373.
\bibitem{zeta}J.S. Dowker and R. Critchley, Phys. Rev. 
\textbf{D 13} (1976) 3224; S.W. Hawking,
Commun. Math. Phys. \textbf{55} (1977) 133. 


\bibitem{Ellis97}
J. Ellis, N.E. Mavromatos and D.V. Nanopoulos,
Int. J. Mod. Phys. {\bf A 12} (1997) 2639.
\bibitem{Mavromatos99}
N.E. Mavromatos and R.J. Szabo, Phys. Rev. {D 59} (1999) 104018.
\bibitem{Ellis98}
J. Ellis, N.E. Mavromatos and D.V. Nanopoulos, {\it
A microscopic Liouville arrow of time}, hep-th/9805120;
{\it Dinamical formation of horizons in recoiling
D branes}, hep-th/0006004.
\bibitem{Polchinski95}
J. Polchinski, Phys. Rev. Lett. {\bf 75} (1995) 4724.

\bibitem{Sen99}
A. Sen, Int. J. Mod. Phys. {\bf A 14} (1999) 4061.
\bibitem{Harvey00}
J.A. Harvey and P. Kraus, JHEP {\bf 0004} (2000) 012.
\bibitem{HKM00}
J.A. Harvey, D. Kutasov and E.J. Martinec, {\it On the relevance of
tachyons}, hep-th/0003101.
\bibitem{Majumder00}
J. Majumder and A. Sen, {\it Vortex pair creation on brane-antibrane
pair via marginal deformation}, hep-th/0003124.
\bibitem{Mello00}
R. de Mello Koch, A. Jevicki, M. Mihailescu and R. Tatar,
{\it Lumps and p-branes in open string theory}, hep-th/0003031.
\bibitem{Moeller00}
M. Moeller, A. Sen and B. Zwiebach, {\it D-branes as tachyons lumps
in string field theory}, hep-th/0005036.
\bibitem{Husain00}
T.Z. Husain and M. Zabzine, {\it Bosonic open strings in a tachyonic
background fields}, hep-th/0005202.
\bibitem{AT1}
O.D. Andreev and A.A. Tseytlin, Nulc. Phys. {\bf B 311} (1988) 205.
\bibitem{AT2}
O.D. Andreev and A.A. Tseytlin, Mod. Phys. Lett. {\bf A 3} (1988) 1349.
\bibitem{Okawa00}
Y. Okawa, Nucl. Phys. {\bf B 566} (2000) 348.
\bibitem{Cornalba99}
L. Cornalba, {\it Corrections to the abelian Born-Infeld action
arising from non-commutative geometry}, hep-th/9912293.
\bibitem{Seiberg99}
N. Seiberg and E. Witten, JHEP {\bf 9909} (1999) 032.
\bibitem{Ho00}
P.-M. Ho and Y.-T. Yeh, {\it Non-commutative D-brane in non-constant
NS-NS B-field}, hep-th/0005159.
\end{thebibliography}
\end{document}